\documentclass[twocolumn,twocolappendix]{aastex63}

\usepackage{amsfonts}
\usepackage{bm}
\usepackage{color}
\usepackage{graphicx}
\usepackage{diagbox}
\usepackage{xspace}
\usepackage{mathtools}

\newcommand{\be}{\begin{equation}}
\newcommand{\ee}{\end{equation}}
\newcommand{\ba}{\begin{eqnarray}}
\newcommand{\ea}{\end{eqnarray}}

\def\DM{\mbox{DM}}
\def\Var{\mbox{Var}}
\def\smallsum{\mathop{\textstyle\sum}\limits}
\def\pcc{pc\,cm$^{-3}$\xspace}

\begin{document}

\title{Mitigating radio frequency interference in CHIME/FRB real-time intensity data}
\shorttitle{Mitigating RFI in CHIME/FRB intensity data}

\author[0000-0001-7694-6650]{Masoud Rafiei-Ravandi}
  \affiliation{Department of Physics, McGill University, 3600 rue University, Montr\'eal, QC H3A 2T8, Canada}
  \affiliation{McGill Space Institute, McGill University, 3550 rue University, Montr\'eal, QC H3A 2A7, Canada}
  \affiliation{Perimeter Institute for Theoretical Physics, 31 Caroline Street N, Waterloo, ON N25 2YL, Canada}
  \affiliation{Department of Physics and Astronomy, University of Waterloo, Waterloo, ON N2L 3G1, Canada}
\author[0000-0002-2088-3125]{Kendrick M.~Smith}
  \affiliation{Perimeter Institute for Theoretical Physics, 31 Caroline Street N, Waterloo, ON N25 2YL, Canada}

\correspondingauthor{Masoud Rafiei-Ravandi}
\email{masoud.rafiei-ravandi@mcgill.ca}

\date{\today}

\begin{abstract}
Extragalactic fast radio bursts (FRBs) are a new class of astrophysical transient with unknown
origins that have become a main focus of radio observatories worldwide. FRBs are highly energetic
($\sim 10^{36}$--$10^{42}$~erg) flashes that last for about a millisecond. Thanks to its broad
bandwidth (400--800 MHz), large field of view ($\sim$200 sq. deg.), and massive data rate (1500 TB
of coherently beamformed data per day), the Canadian Hydrogen Intensity Mapping Experiment / Fast
Radio Burst (CHIME/FRB) project has increased the total number of discovered FRBs by over a factor
10 in 3 yr of operation. CHIME/FRB observations are hampered by the constant exposure to radio
frequency interference (RFI) from artificial devices (e.g., cellular phones, aircraft), resulting
in $\sim$20\% loss of bandwidth. In this work, we describe our novel technique for mitigating RFI
in CHIME/FRB real-time intensity data. We mitigate RFI through a sequence of iterative operations,
which mask out statistical outliers from frequency-channelized intensity data that have been
effectively high-pass filtered. Keeping false-positive and false-negative rates at very low levels,
our approach is useful for any high-performance surveys of radio transients in the future.
\end{abstract}


\section{introduction}
\label{sec:introduction}

The search for extragalactic fast radio bursts (FRBs) has become a central undertaking
for radio telescopes worldwide. FRBs are brief ($\sim$millisecond) flashes of radio signal with
unknown origins. To date, over 750 FRBs have been discovered \citep[see, e.g., CHIME/FRB Catalog 1
in][]{Collaboration:2021wz}.\footnote{Catalogs of known FRBs
are available on \url{https://www.herta-experiment.org/frbstats} \citep{Spanakis-Misirlis:2021ta}
and \url{https://www.wis-tns.org} \citep{2020TNSAN.160....1P}.} This includes 25
repeating sources that emit multiple bursts sporadically over long ($\sim$days to months)
time intervals, and about 20 (repeating and nonrepeating) sources that are localized to
host galaxies \citep[for a recent review, see][]{Petroff:2022wo}.

The FRB arrival time $t$ at the observing frequency $\nu$ is delayed proportional to $\nu^{-2}$, 
owing to the dispersion of radio waves in the intervening cold plasma along the line of sight.
The delay is measured by the dispersion measure (DM), which sets the proportionality constant
in the dispersion relation. FRBs are found by integrating the total intensity $I(t,\nu)$ across
frequency along different dispersion profiles and searching for sharp peaks in the time profiles
in the ($t, \DM$) space.

Most FRBs have been discovered by the Canadian Hydrogen Intensity Mapping Experiment /
Fast Radio Burst (CHIME/FRB) instrument \citep{Collaboration:2018aa}, which is equipped with
an incoherent dedispersion search engine (out to a maximum DM of 13,000 \pcc) operating constantly
between 400 and 800 MHz. CHIME/FRB is one of the four digital backends that are connected to the
CHIME telescope, which has a large field of view ($\sim$200 deg$^2$). The CHIME telescope and its
digital backends \citep{CHIME:2022} are located at the Dominion Radio Astrophysical Observatory
(DRAO), which is in a radio-quiet area in the Okanagan Valley, British Columbia, Canada. Despite
being regulated by local and federal governments, the CHIME radio bandwidth is exposed to a broad
range of anthropogenic signals from various sources nearby (e.g., cellular phones) and far away
(e.g., aircraft and satellites) that interfere with daily observations of astrophysical phenomena.
If not mitigated properly, even a very short ($\sim$10\,ms) blip of radio frequency interference
(RFI) could trigger thousands of false positives in the CHIME/FRB instrument.

RFI can be very unpredictable in the $(t,\nu)$ space. Several RFI mitigation
techniques have been effective in the past. Most notably, impulsive RFI could be masked
through an iterative threshold-based scheme \citep[see, e.g., the SumThreshold
algorithm in][]{SumThreshold, SumThresholdLOFAR} or a statistical estimator such as the
spectral kurtosis \citep[see, e.g.,][]{SK_general1, SK_general2, CHIME_RFI}. Periodic RFI could
be detected more robustly in the harmonic space of intensity \citep[see, e.g.,][]{FFT_RFI}.
In addition, large-scale RFI variations could be removed by fitting a baseline
\citep[see, e.g.,][]{Eatough:2009vn, FAST}. Furthermore, general RFI patterns could be
flagged through machine-learning algorithms \citep[see, e.g.,][]{CNN, CNN2, CNN_FAST}.
In this work, we describe our strategy for mitigating RFI in the CHIME/FRB data. CHIME/FRB science
requirements combined with its data rate (1500 TB/day) introduced computational challenges
that led to the design and development of extremely optimized software from scratch.

As a brief overview, CHIME/FRB masks RFI through a long sequence of iterative operations in
a real-time pipeline. These operations have been fine-tuned in order to account for the computational
cost while simultaneously minimizing the false-positive and false-negative rates. The pipeline computes
a time-varying RFI mask for the frequency-channelized intensity data, so that RFI can be suppressed
prior to the dedispersion transform. This technique is the first step in our RFI mitigation and is
the only one that works directly with the intensity data to lower the number of false positives
from $\sim10^5$ to a few per day, while detecting FRBs at an unprecedented rate
\citep[][]{Collaboration:2021wz}. Despite being designed for a specific instrument, our algorithms
could in principle be applied to other real-time searches of radio transients such as pulsar surveys.

The paper is organized as follows. We present an overview of the CHIME/FRB software pipeline in
\S\ref{sec:chimefrb_pipeline}. In \S\ref{sec:rfi_transform_chains}, we present a detailed
account of our RFI mitigation strategy. We discuss results, unsolved problems, and potential
opportunities for improvement in \S\ref{sec:discussion}. Throughout, we adopt the CHIME/FRB
data as a running example.

\section{CHIME/FRB pipeline}
\label{sec:chimefrb_pipeline}

\begin{figure*}
\centerline{
        \includegraphics[width=15.017676767676768cm]{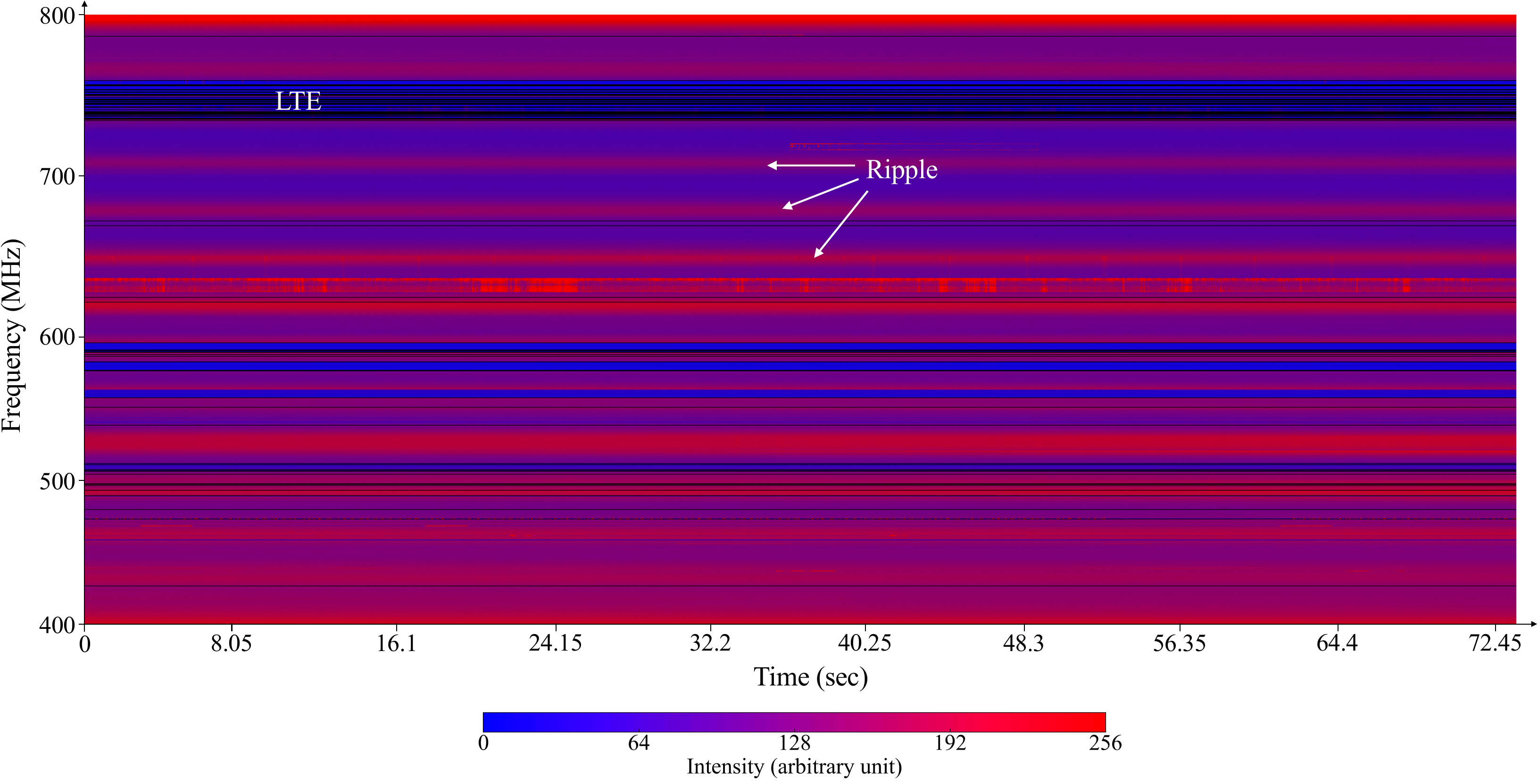}
}
\vspace{0.1cm}
\centerline{
        \includegraphics[width=15.017676767676768cm]{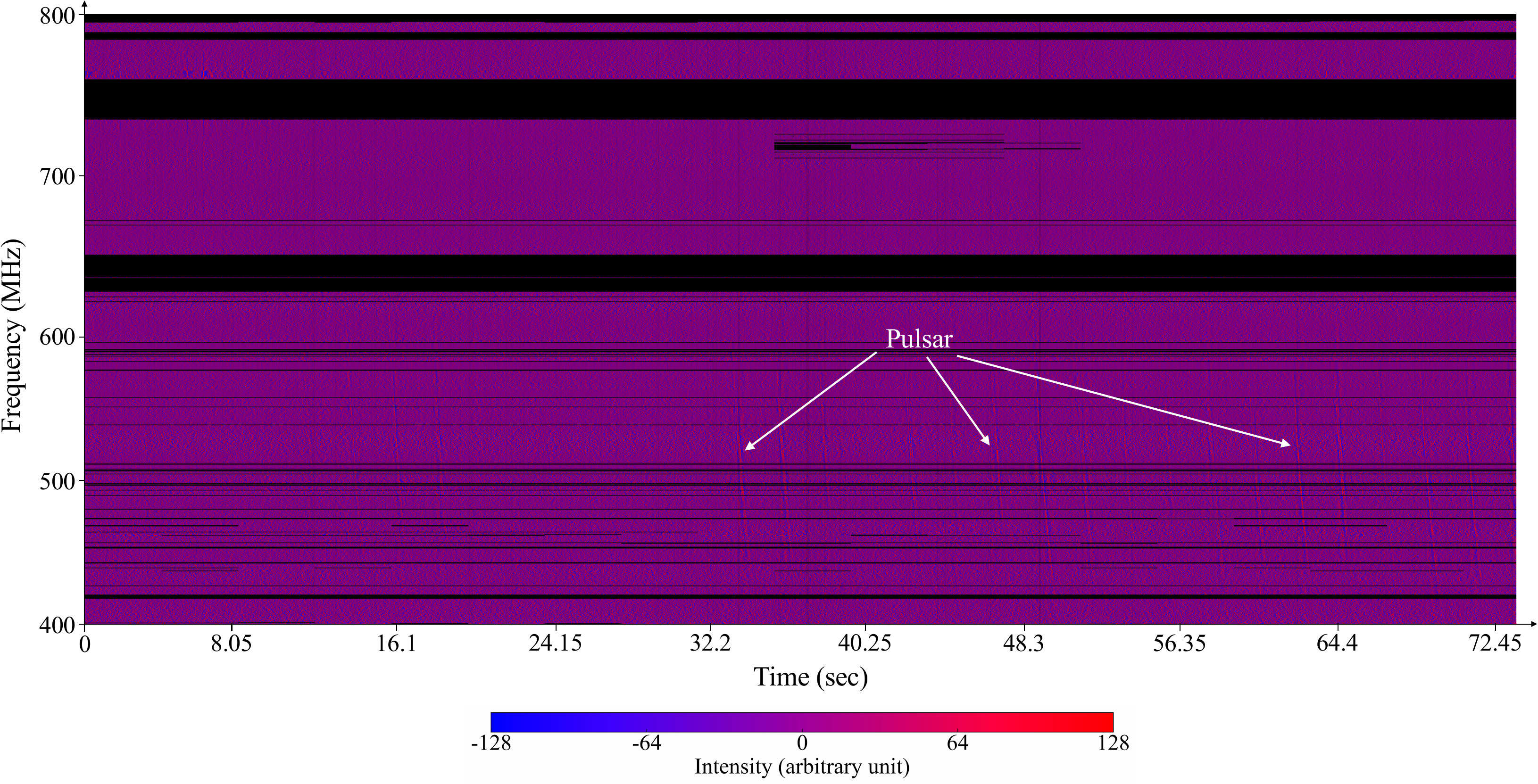}
}
\vspace{0.1cm}
\centerline{
        \hspace*{-0.18cm}\includegraphics[width=15.2cm]{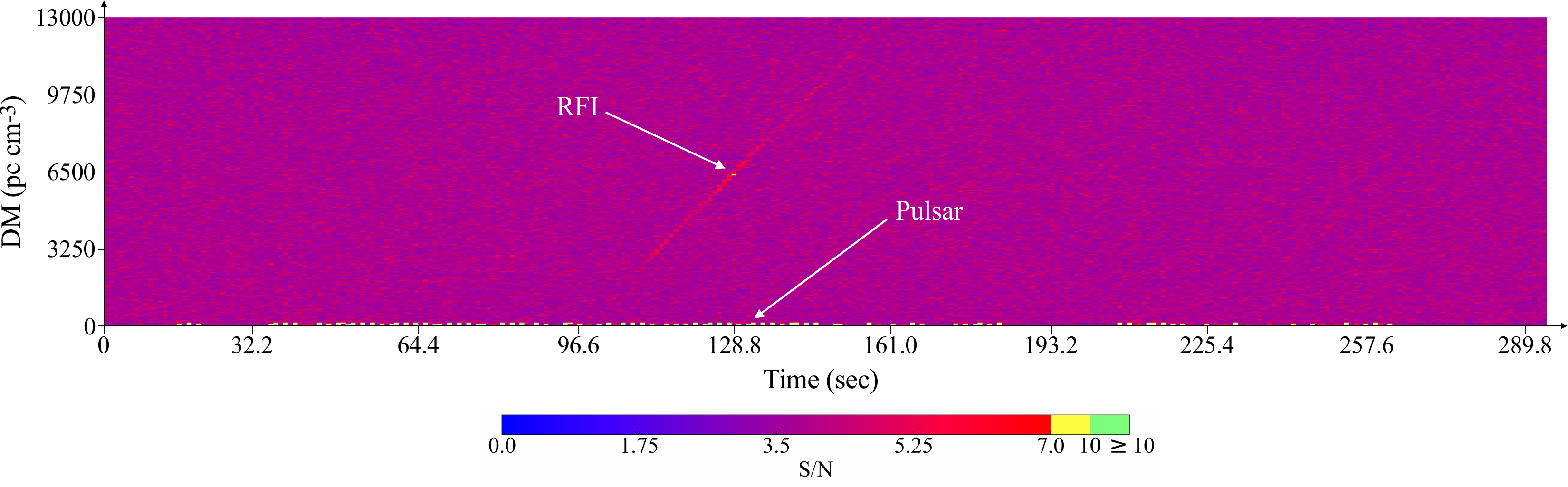}
}
\caption{CHIME/FRB L1 data for a single beam, containing a pulsar transit along with RFI.
         {\em Top panel:} raw intensity with persistent RFI between 600 and 650\,MHz. The
         relatively low-intensity region centered at $\sim$750\,MHz corresponds to the long-term
         evolution (LTE) band, which is usually active. Multiple scatterings of radio waves off
         the feed structure cause the systematic ``ripple'' effect with a period of
         $\approx$30\,MHz across the bandwidth \citep[][]{CHIME:2022}. Black regions indicate
         missing data. {\em Middle panel:} detrended intensity, including the missing data and
         RFI mask (black regions) prior to the dedispersion transform (see
         Figure~\ref{fig:l1_diagram_high_level}). Dispersed sweeps ($\propto\nu^{-2}$) across
         the bandwidth, particularly visible in the 600--400\,MHz range at $\approx$2.2 second
         intervals, are signals from a nearby pulsar.
         {\em Bottom panel:} dedispersion space of events, including the pulsar (green and yellow
         dots at DM $\lesssim 100$ \pcc). The RFI event at ($t \approx 128.8$\,s, DM $\approx
         6500$ \pcc) has a low signal-to-noise ratio ($7 \le {\rm S/N} < 10$) and is below the
         detection threshold.}
\label{fig:waterfall}
\end{figure*}

The CHIME/FRB instrument \citep{Collaboration:2018aa} has multiple processing levels
(denoted by L1, L2, etc.) that have been running constantly except during upgrade cycles and
emergency shutdowns since mid-2018 \citep[for a full discussion of telescope up time, see][]{CHIME:2022}.
In L0 \citep[the CHIME FX correlator, containing 256 processing nodes; also see][]{CHIME:2022},
frequency-channelized data are coherently arranged in an array of $4\times256$ digitally formed beams.
In L1 (the real-time FRB search engine, running on 128 processing nodes, each with two 10-core Intel
E5-2630v4 CPUs and 128 GB of RAM), beamformed packets are assembled into a time series of
total intensity samples. Then, RFI is mitigated by generating a mask (zeros and ones), which is
subsequently used for detrending (or effectively high-pass filtering) and dedispersing the intensity
(see Figure~\ref{fig:waterfall}). Once the intensity is dedispersed, L1b groups and ranks
astrophysical vs. RFI-triggered events based on their signal-to-noise ratio (S/N), arrival time, DM,
and sky location. Next, L2/L3 labels sources ``Galactic'' or ``extragalactic'' and depending on this
and on, e.g., S/N, decides what action to take (e.g., whether to store buffered intensity data).
Finally, metadata are stored in a database (L4).

We define a ``false positive'' as an event with nonastrophysical origin (e.g., RFI, thermal noise,
and faulty calibration) and S/N\,$\geq$\,10 that passes through L1--L4 and is classified as an
astrophysical candidate. We emphasize that the main CHIME/FRB search is carried out over the
dedispersion space. Resembling FRBs by randomly lining up as $\nu^{-2}$ in the intensity space,
RFI has been by far the largest contributing factor to the CHIME/FRB false positives, which could
easily overwhelm L2--L4 pipelines. In addition, unmasked RFI could reduce the true-positive rate.
The L1 system computes a running estimate of the variance of intensity, and unmasked RFI could
contaminate this estimate, lowering the overall system sensitivity, severely impacting the
true-positive rate for various selection criteria. As such, we mitigate false positives in all levels
of the intensity pipeline: L1 masks raw intensity data without incorporating beam information,
while L1b--L3 employ machine-learning classifiers to discriminate between true and false positives
after sky localization utilizing ancillary information about the system performance and environment.
This paper focuses on the RFI mitigation prior to the dedispersion transform in L1.

\begin{figure*}
\centerline{
    \includegraphics[width=16.8cm]{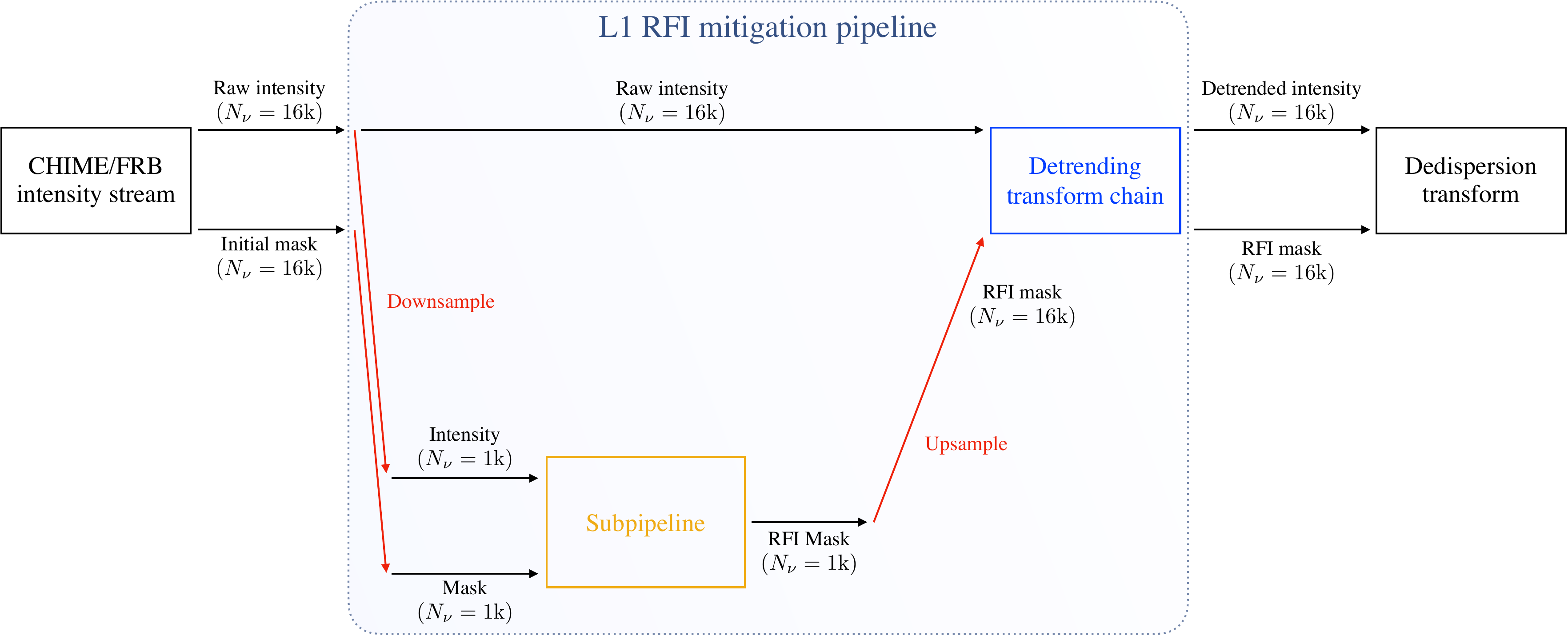}
}
    \caption{CHIME/FRB L1 pipeline for processing frequency-channelized intensity data
    in real time. L1 mitigates RFI for 16k frequency channels ($N_\nu={\rm 16k}$) by
    constantly generating a downsampled mask through a long sequence of operations
    inside the ``subpipeline,'' which reduces the overall computational cost by a factor
    $\approx$16 (see Figure~\ref{fig:l1_diagram_subpipeline}). The RFI mask is upsampled
    back to 16k frequency channels for detrending and dedispersing the raw intensity.}
\label{fig:l1_diagram_high_level}
\end{figure*}

The L1 software package contains multiple object-oriented code libraries with distinct
purposes. For instance, ``{\tt \detokenize{ch_frb_io}}''\footnote{{\tt \detokenize{ch_frb_io}}:
\url{https://github.com/CHIMEFRB/ch_frb_io}} assembles L0 data packets into a stream of 1\,s
chunks of masked intensity data,\footnote{Intensity values are floating-point numbers between 0
and 256. Initial masks contain zeros and ones that correspond to bad and good intensity data,
respectively.} which are then fed into ``{\tt \detokenize{rf_pipelines}}''\footnote{{\tt
\detokenize{rf_pipelines}}: \url{https://github.com/kmsmith137/rf_pipelines}} for RFI mitigation.
Figure~\ref{fig:l1_diagram_high_level} shows the high-level logic of the RFI mitigation pipeline
as configured and implemented for CHIME/FRB.

Each box in the diagram represents an operation (or ``transform") whose input and output arrays are
represented as arrows. All arrays are 2D arrays indexed by ($t, \nu$), which are processed in 4\,s
chunks (see \S\ref{sec:rfi_transform_chains}). The ``intensity stream'' continuously emits arrays of
the intensity $I_{t\nu}$ and mask $M_{t\nu}$ that are downsampled from 16k to 1k frequency channels
as follows:
\ba
I_{t\nu} &=& \frac{\smallsum_{n=0}^{15}{M_{t,16\nu+n} \, I_{t,16\nu+n}}}{\smallsum_{n=0}^{15}{M_{t,16\nu+n}}} \label{eq:downsampling_i} \\
M_{t\nu} &=& \frac{1}{16}\smallsum_{n=0}^{15}{M_{t,16\nu+n}} \, , \label{eq:downsampling_w}
\ea
where the masks are treated as nonbinary weights. Next, downsampled data are processed inside the
``subpipeline'' in order to update the mask, which is then upsampled from 1k to 16k frequency channels
for mitigating RFI in the raw intensity data. We upsample arrays by copying and tiling elements
along the downsampled axis (e.g., upsampling the 1D array $[0,0.5,1]$ by a factor 2 yields
$[0,0,0.5,0.5,1,1]$). This down/upsampling logic is essential for minimizing the computational cost while
maintaining a robust yet coarse-grained estimate of time-varying RFI patterns across the CHIME/FRB bandwidth
at 1\,ms time resolution. In \S\ref{sec:rfi_transform_chains}, we take a deep dive into the core of our RFI
mitigation algorithm inside the ``subpipeline.''

Using the upsampled mask, we detrend and dedisperse the \emph{raw} intensity with 16k frequency channels.
Thus, the dedispersion transform continuously receives a set of (detrended intensity, RFI mask) arrays
with 16k frequency channels as input. Then, the dedispersion transform replaces masked
regions\footnote{The dedispersion transform can be configured to expand the boundaries of masks in order
to suppress potentially unmasked residuals from highly active RFI in neighboring frequency and time samples.
This is the last line of defense against RFI-triggered false positives in L1.} of intensity arrays by an
estimate of the running variance based on a white noise model for the input intensity values and a slowly
varying function for the variance of unmasked intensity along the time axis. In addition, the running variance
is adopted for normalizing dedispersed events to S/N. Finally, the dedispersion transform outputs a 4D array
(DM, scattering measure, spectral index, time) of coarse-grained trigger statistics, where the coarse-graining
is done by taking the maximum over time and DM.

The entire chain of L1 operations is fixed in the CHIME/FRB instrument; a slight modification in the
algorithm requires a system-wide reconfiguration and restart.\footnote{CHIME/FRB L1 configuration files
are available in {\tt \detokenize{ch_frb_rfi}}: \url{https://github.com/mrafieir/ch_frb_rfi}} Although
very rare, such reconfigurations are necessary when RFI starts to show up in a newly allocated radio
band, e.g., owing to the 5G network. To that end, we alert system operators\footnote{The instrument is
monitored daily by members of the CHIME/FRB collaboration.} to save several hours of intensity data
during active RFI. Using the saved data, we attempt to reproduce the culprit for a close visual
inspection. We can often find a simple hotfix that will suppress the new RFI effectively
(see \S\ref{ssec:auxiliary_transforms}).

\section{RFI transform chains}
\label{sec:rfi_transform_chains}

The ``subpipeline'' (see Figure~\ref{fig:l1_diagram_high_level}) contains a sequence of over 100
transforms for masking RFI. This long chain of operations is constructed mainly by iterating over
the clipping and detrending transforms, which are described in \S\ref{ssec:clipping_transform_chain}
and \S\ref{ssec:detrending_transform_chain}, respectively. We took the iterative approach based
on the observation that RFI contaminates the probability distribution function (PDF) of intensity
(and its higher moments) through statistical outliers (e.g., 5$\sigma$ deviations from the mean of
intensity in a specific frequency channel) that can be masked incrementally.

While our clipping and detrending transforms can generally recognize and suppress RFI through
the upsampled mask, they can also leave some residuals that induce false positives after the
final round of detrenders. For example, unmasked narrowband RFI could cause a series of
ripples in the detrended intensity along the frequency axis. Under the right circumstances, these
ripples could induce false positives downstream. Therefore, we need to account for such
residual-induced false positives as well as RFI-triggered false positives throughout the L1
pipeline (see, e.g., Figure~\ref{fig:waterfall_fp}). In addition, we require an overall latency of
$\sim$10\,s for reporting events in real time. This is to facilitate detection of counterparts
using other telescopes, achieved through automated real-time alerts \citep{Zwaniga:2021}.
Also, our baseband (voltage) recording system \citep{Michilli:2021tm} imposes
an upper bound (4\,s) on the size of intensity chunks in the L1 ring buffers.

\begin{figure*}
\centerline{
    \includegraphics[width=12.860484347312465cm]{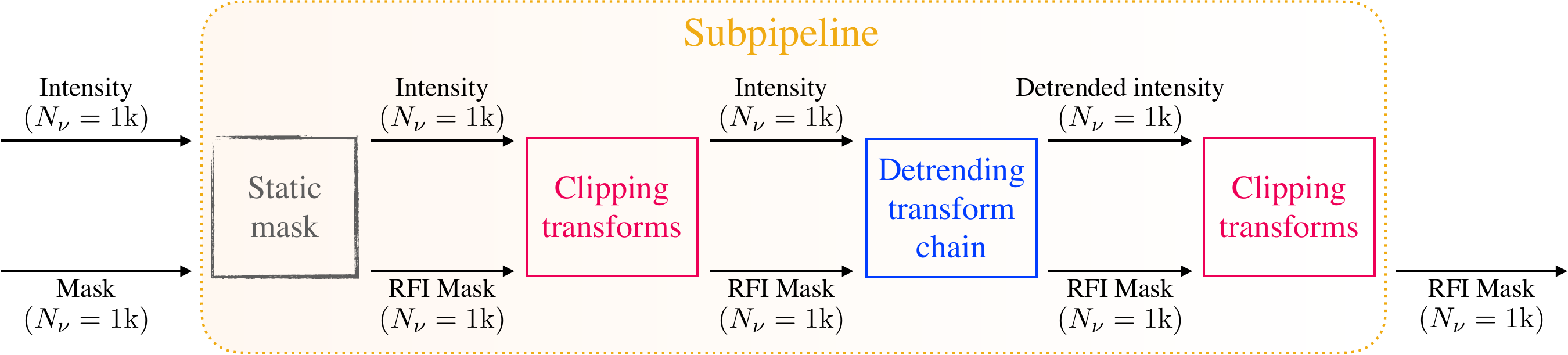}
}
\vspace{0.3cm}
\centerline{
    \includegraphics[width=7.819492025989368cm]{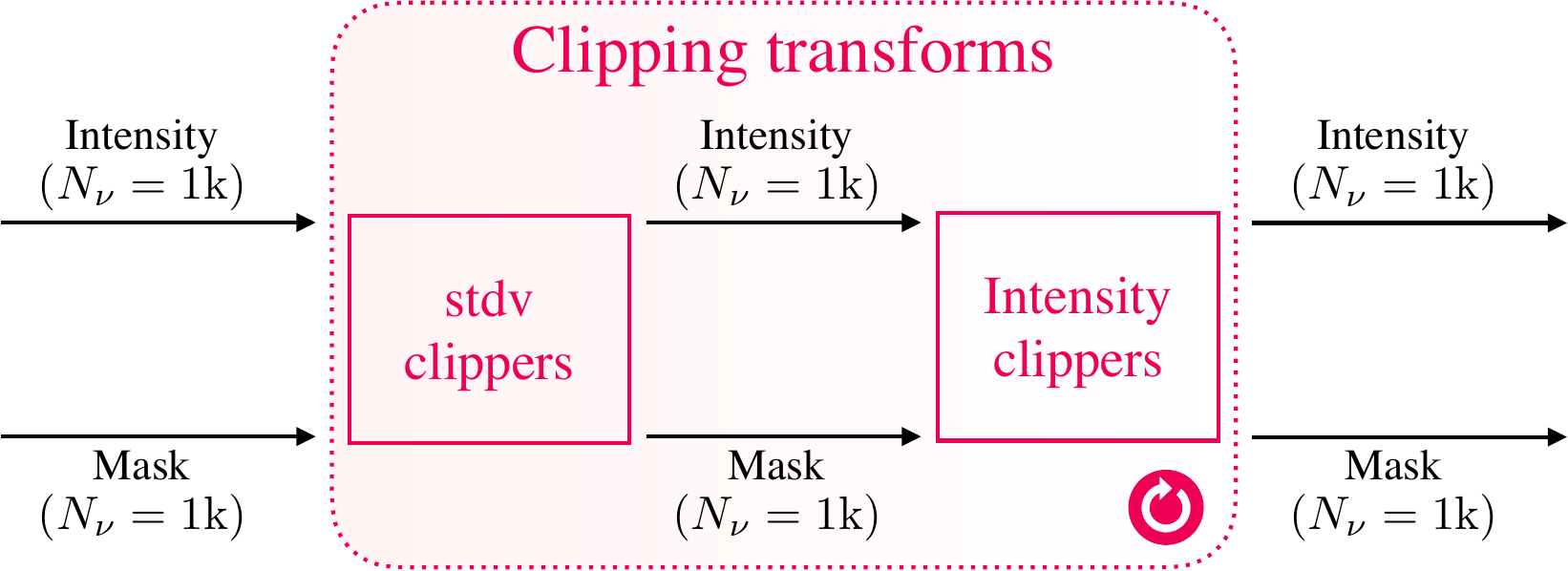}
}
\vspace{0.3cm}
\centerline{
    \includegraphics[width=7.819492025989368cm]{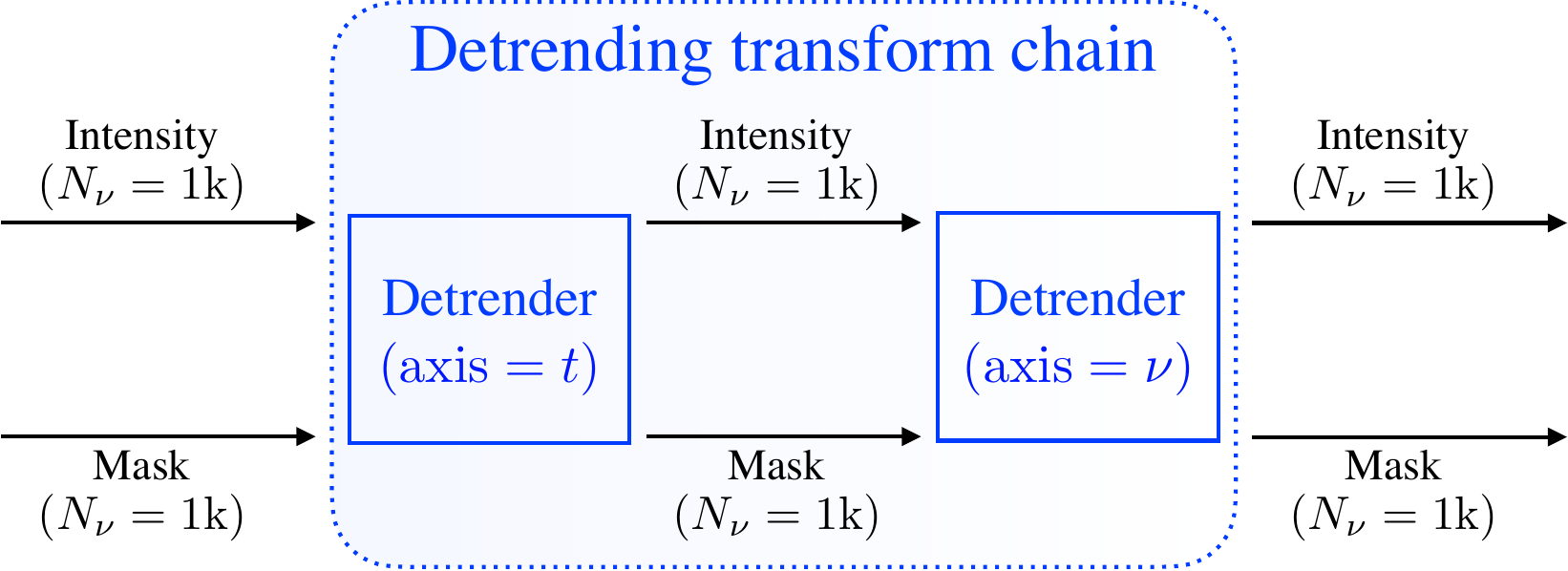}
}
    \caption{CHIME/FRB L1 subpipeline for generating dynamic RFI masks over intensity
    chunks with 1k frequency channels.
    \emph{Top panel:} high-level diagram of the subpipeline, including the initial static
    mask, clipping and detrending transform chains in a specific order as described
    in \S\ref{sec:rfi_transform_chains}.
    \emph{Middle panel:} clipping transform chain that is iterated six times (circular arrow).
    \emph{Bottom panel:} detrending transform chain, which is also executed for raw intensity
    data with 16k frequency channels (see Figure~\ref{fig:l1_diagram_high_level}).
    }
\label{fig:l1_diagram_subpipeline}
\end{figure*}

Figure~\ref{fig:l1_diagram_subpipeline} illustrates the CHIME/FRB RFI transform chain nested
inside the subpipeline. 
The simplest form of RFI transforms is a static mask, which constantly eliminates a set of known
frequency channels exhibiting intense RFI. We place the static mask at the very beginning of our
chain in order to minimize residuals, which can easily distort the intensity PDF.

Following the static mask, clipping transforms or ``clippers'' take the streaming chunk of intensity
data and mask out statistical outliers dynamically. These outliers could be deviations from the mean
or standard deviation (stdv) of the intensity PDF along an arbitrary axis (i.e., time and/or frequency).
Then, a sequence of detrending transforms or ``detrenders'' remove large-scale intensity variations
(along time and frequency) from the data. Next, a second round of clipping transforms is applied to
the modified intensity PDF. We note that both the mask and intensity data stream are updated throughout
this process inside the subpipeline. However, only the mask is upsampled and passed to the last round
of detrenders prior to the dedispersion over 16k frequency channels. The last round of detrenders
is required for the CHIME/FRB dedispersion transform, which assumes a zero baseline for the incoming
intensity data stream. Finally, the upsampled mask ($\sim$20\% loss of bandwidth on average) and
detrended intensity data are passed to the dedispersion transform.

Ideally, the specific configuration of RFI transform chains would be determined through
a global minimization problem, which would aim at minimizing the computational cost along
with the false-positive and false-negative rates. In actuality, we configured the CHIME/FRB
RFI transform chain empirically; although not necessarily optimal, we configured transforms
through trial and error using hundreds of hours of intensity acquisitions.\footnote{Other than with 
the real-time pipeline, L1 can be run in an offline mode with saved intensity data.} The
remainder of this section contains a detailed account of the RFI transform chains.

\subsection{Clipping transform chain}
\label{ssec:clipping_transform_chain}

CHIME/FRB beamformed intensity data are constructed by summing squared
voltages $\langle V_a(\nu) V_b^*(\nu') \rangle$ from the CHIME antennas $a$ and $b$
\citep[see, e.g.,][]{beamforming}. In the absence of RFI, we expect the raw voltages
to be Gaussian distributed \citep{Fridman:2001wy} and hence the intensity to be
$\chi^2$ distributed. In CHIME/FRB, we developed two different types of transforms
for masking statistical outliers in the intensity data: intensity and standard deviation
(stdv) clipping transforms.

Let $I_{t\nu}$ represent the 2D array of possibly downsampled {\em masked}
intensity, which is a function of time ($t$) and frequency ($\nu$). Then, the
intensity clipping transform updates the mask $M_{t\nu}$ by zeroing out elements that satisfy
\be
\left| I_{t\nu} - \bar{I}_{t\nu} \right| \ge \left(\sigma_c \Var(I)_{t\nu}^{1/2}\right) ,
\label{eq:intensity_clipper}
\ee
where the threshold level $\sigma_c$ is a configurable constant, and the mean intensity
array $\bar{I}_{t\nu}$ and variance array $\Var(I)_{t\nu}$ are computed based
on the initial intensity $I_{t\nu}$ along a specific axis in time and/or frequency
in two stages as follows:
\be
I_{t\nu}
\xrightarrow[{\rm internally}]{\rm clip} \left[\bar{I}_{t'\nu'}, \Var(I)_{t'\nu'}\right]
\xrightarrow{{\rm reshape}} \left[\bar{I}_{t\nu}, \Var(I)_{t\nu}\right] ,
\label{eq:intensity_clipper_nested}
\ee
where the indices $t'$ and $\nu'$ refer to arrays with a reduced length,
e.g., computing the mean along the time axis of an array with shape
($N_t, N_\nu$)=(4k,\,16k) results in a new array with shape
($N_{t'}, N_{\nu'}$)=(1,\,16k). In Equation~(\ref{eq:intensity_clipper_nested}),
the first stage is the computation of the {\em masked} mean and variance along
an axis:
\ba
\bar{I}_{t'\nu'} &=& \frac{\smallsum_{m=0}^{D_t-1}\,
                           \smallsum_{n=0}^{D_\nu-1} {M_{xy} \, I_{xy}}}
                           {\smallsum_{m=0}^{D_t-1}\, \smallsum_{n=0}^{D_\nu-1} {M_{xy}}} \label{eq:mean_ibar} \\
\Var(I)_{t'\nu'} &=& \frac{\smallsum_{m=0}^{D_t-1}\,
                           \smallsum_{n=0}^{D_\nu-1} M_{xy} {\left(I_{xy} - \bar{I}_{xy}\right)^2}}
                           {\smallsum_{m=0}^{D_t-1}\, \smallsum_{n=0}^{D_\nu-1} {M_{xy}}} , \label{eq:var_i}
\ea
where the indices $x\equiv t D_t + m$ and $y\equiv \nu D_\nu + n$ are defined with the
downsampling factors $D_t \equiv N_{t'} / N_t$ and $D_\nu \equiv N_{\nu'} / N_\nu$ along
the time and frequency axis, respectively. Here, the possibly downsampled mask $M_{xy}$ is
treated as nonbinary weights. In addition,
an intensity clipping transform can have an internal iteration that temporarily masks out
intensity outliers $n_{c,{\rm in}}$ times prior to the final computation of mean and variance.
We configure the internal iteration by the threshold parameter $\sigma_{c,{\rm in}}$. Whether
internal or external, every clipping iteration helps improve the RFI mask by recognizing more
statistical outliers and hence reshaping the masked intensity PDF to a robust $\chi^2$
distribution in real time. In the second stage, we reshape these {\em masked} arrays by
copying and tiling elements along the reduced axis. This recovers the initial shape of input
intensity data, enabling the final element-wise operation in Equation~(\ref{eq:intensity_clipper}).

Likewise, the stdv clipping transform masks RFI along the time {\em or} frequency axis using
the following criterion:
\be
\left| J_{t'\nu'} - \overline{J_{t'\nu'}} \right| \ge
\left( \sigma_c \Var(J_{t'\nu'})^{1/2} \right) ,
\label{eq:stdv_clipper}
\ee
where the masked array $J_{t'\nu'} \equiv \Var(I)_{t'\nu'}$ with {\em no internal clipping},
and the mean of variance $\overline{J_{t'\nu'}}$ and variance of variance $\Var(J_{t'\nu'})$
are scalar. Unlike the intensity clippers, the stdv clippers ``reshape'' the running mask to
match its shape ($N_{t'}, N_{\nu'}$) with the initial intensity $I_{t\nu}$ only {\em after}
executing Equation~(\ref{eq:stdv_clipper}).

Table \ref{tab:clipping_transforms} lists the CHIME/FRB clipping transform chain, which
is iterated $n_c$ times before and after the intensity detrenders, i.e., the clipping transform
chain is a nested loop inside an outer loop over detrenders. The first three stdv clippers are along
the time axis, and have a 3$\sigma$ threshold with no down/upsampling factors. Then, two stdv clippers
are applied along the frequency axis, with a 3$\sigma$ threshold and no down/upsampling factors.
Next, two intensity clippers are applied at a 5$\sigma$ threshold. Finally, two intensity clippers are
applied with downsampling factors of 16 and 2 along the time and frequency axis, respectively.
The downsampling logic helps expand the mask and hence prevent intensity leakage over RFI-contaminated regions.

\begin{table}
\begin{center}
\begin{tabular}{ccccccccc}
\hline\hline
    $i$ &
    type &
    $\sigma_c$ &
    $\sigma_{c,{\rm in}}$ &
    $n_{c,{\rm in}}$ &
    axis &
    $N_t$ &
    $D_t$ &
    $D_\nu$ \\
    \hline
    1 & stdv & 3 & $-$ & $-$ & $t$ & 4096 & 1 & 1\\ 
    2 & stdv & 3 & $-$ & $-$ & $t$ & 4096 & 1 & 1\\
    3 & stdv & 3 & $-$ & $-$ & $t$ & 4096 & 1 & 1\\
    4 & stdv & 3 & $-$ & $-$ & $\nu$ & 4096 & 1 & 1\\
    5 & stdv & 3 & $-$ & $-$ & $\nu$ & 4096 & 1 & 1\\
    6 & intensity & 5 & 5 & 9 & $\nu$ & 4096 & 1 & 1\\
    7 & intensity & 5 & 5 & 9 & $t$ & 4096 & 1 & 1\\
    8 & intensity & 5 & 3 & 9 & $(t, \nu)$ & 4096 & 16 & 2\\
    9 & intensity & 5 & 3 & 9 & $\nu$ & 4096 & 16 & 2\\ \hline\hline
\end{tabular}
\end{center}
    \caption{Clipping transform chain for dynamically masking RFI in the CHIME/FRB intensity stream.
    The index $i$ specifies the position of each transform in the chain. The intensity and
    standard deviation (stdv) transforms clip statistical outliers at the threshold level
    $\sigma_c$ from the mean (see Equations~\ref{eq:intensity_clipper} and \ref{eq:stdv_clipper}).
    In addition, the intensity clipping transform has an internal iteration
    which allows for temporarily masking outliers at the $\sigma_{c,{\rm in}}$ level prior to the final
    clipping operation. The number of internal iterations is set by the parameter $n_{c,{\rm in}}$.
    $N_t$ is the total number of 1\,ms time samples in an intensity chunk that each transform processes
    in real time. In each transform, the incoming chunks are downsampled (and consequently final
    coarse-grained masks are upsampled) by the two factors $D_t$ and $D_\nu$ along the time and
    frequency axis, respectively (see, e.g., Equations~\ref{eq:mean_ibar} and \ref{eq:var_i}).
    Operations along the axis $(t, \nu)$ are carried out over the 2D array of intensity values.}
\label{tab:clipping_transforms}
\end{table}

Since RFI is a non-Gaussian process \citep{Fridman:2001wy,Mirhosseini:2020tb}, we expect RFI-triggered
false positives to appear more distinctly at higher statistical moments of the intensity. Therefore,
we place the stdv transforms at the very beginning of the clipping transform chain. Additionally, we
refrain from downsampling the intensity in the beginning, since coarse-grained masks lack precision in
containing only regions with RFI. In the case of intensity clippers, internal iterations are effective
in decontaminating PDFs at a low computational cost. On the other hand, RFI residuals can hide behind
internal masks and hence contaminate the stream if $\sigma_{c,{\rm in}}$ and $n_{c,{\rm in}}$ are not
set properly. Finally, we mask typically along the time axis first, since RFI tends to arrive in constant
frequency bands. Furthermore, we maximize the chunk size $N_t$, which is constrained mainly by the latency
requirement for the CHIME/FRB baseband recording system. We find empirically that by iterating the
clipping transform chain (Table \ref{tab:clipping_transforms}) six times before and after a chain of
detrenders (see \S\ref{ssec:detrending_transform_chain}), we are able to sufficiently suppress
RFI while detecting real pulsars in CHIME/FRB intensity acquisitions.\footnote{We initially designed
the RFI transform chain through the analysis of incoherent beam data (400--800\,MHz, $N_\nu$=1k) from
the CHIME Pathfinder \citep[][]{Bandura:2014uc} and single-beam data (400--800\,MHz, $N_\nu$=16k) from
the Galt 26\,m telescope. These instruments share the same RFI environment with CHIME/FRB at DRAO. The
current configuration is based on seven different production-level transform chains that were designed
to either mitigate new sources of RFI or adapt to other pipeline requirements (e.g., constraining the buffer
size $N_t$ for the baseband recording system) sporadically since the first CHIME/FRB commissioning in
mid-2018.}

\subsection{Detrending transform chain}
\label{ssec:detrending_transform_chain}

The CHIME/FRB intensity stream exhibits large-scale variations from RFI, forward gains,
and digital beamforming \citep{Ng:2017ww} as functions of the time, frequency, and sky
location. Such baseline variations distort the intensity PDF, causing the clipping and
dedispersion transforms to fail. We mitigate this effect by introducing a chain of detrending
transforms in the pipeline. Detrending transforms remove a best-fit model with the polynomial
degree $\alpha$ from the masked intensity array $I_{t\nu}$ along an axis \citep[see,
e.g.,][]{Eatough:2009vn}. This real operation is a computationally low-cost alternative
to a high-pass filter in the harmonic space of intensity.

Table~\ref{tab:detrending_transforms} lists the CHIME/FRB detrending transform
chain, which contains two types of transforms: the polynomial (based on Legendre
polynomials) and spline detrenders, both of which make use of the Cholesky decomposition
for efficient computation of the best-fit coefficients. We designed the main RFI transform
chain in the subpipeline such that detrenders are sandwiched between clipping transforms (see
Figure \ref{fig:l1_diagram_subpipeline}). In other words, the detrending transform chain is an
outer loop with $n_d$ iterations, each of which contains $n_c$ inner iterations over the clipping
transform chain. We emphasize that subpipeline detrenders do not affect the intensity
stream, which is fed into the dedispersion transform (see Figure~\ref{fig:l1_diagram_high_level}).
Subpipeline detrenders work in tandem with clipping transforms in order to generate a mask
that mitigates RFI contamination in the detrended downsampled intensity chunks.

\begin{table}
\begin{center}
\begin{tabular}{c@{\hskip 0.8cm}c@{\hskip 0.8cm}c@{\hskip 0.8cm}c@{\hskip 0.8cm}c}
\hline\hline
    $j$ &
    type &
    $\alpha$ &
    axis &
    $N_t$ \\
    \hline
    1 & polynomial & 4 & $t$ & 4096\\
    2 & spline & 12 & $\nu$ & 4096\\ \hline\hline
\end{tabular}
\end{center}
    \caption{Detrending transform chain for the CHIME/FRB intensity stream.
    The index $j$ specifies the position of each transform in the chain.
    Each transform fits a model with the polynomial degree $\alpha$ to intensity
    arrays of width $N_t$ (number of 1\,ms time samples) along one ``axis''.
    Then, the model is subtracted from data in order to account for large-scale
    variations due to RFI and instrumental systematics.}
\label{tab:detrending_transforms}
\end{table}

The CHIME/FRB subpipeline currently has only a single iteration ($n_d=1$) of the detrending transform
chain. We detrend the \emph{raw} intensity after generating the RFI mask; using the upsampled mask with
16k frequency channels, we apply the same detrending transform chain to the raw intensity ($N_\nu=16$k)
prior to the dedispersion transform, which assumes input data with a zero mean. In addition to modifying the
intensity values, detrenders can modify the running mask (e.g., by masking an entire chunk) in corner
cases where numerical instabilities, e.g., owing to highly active RFI, cause a failure in converging
on a robust solution. Therefore, the upsampled RFI mask is subject to further modification in the last
round of detrenders between the subpipeline and dedispersion transform
(see Figure~\ref{fig:l1_diagram_high_level}).

\subsection{Auxiliary transforms}
\label{ssec:auxiliary_transforms}

\begin{figure*}
\centerline{
        \includegraphics[width=5.597150259067358cm]{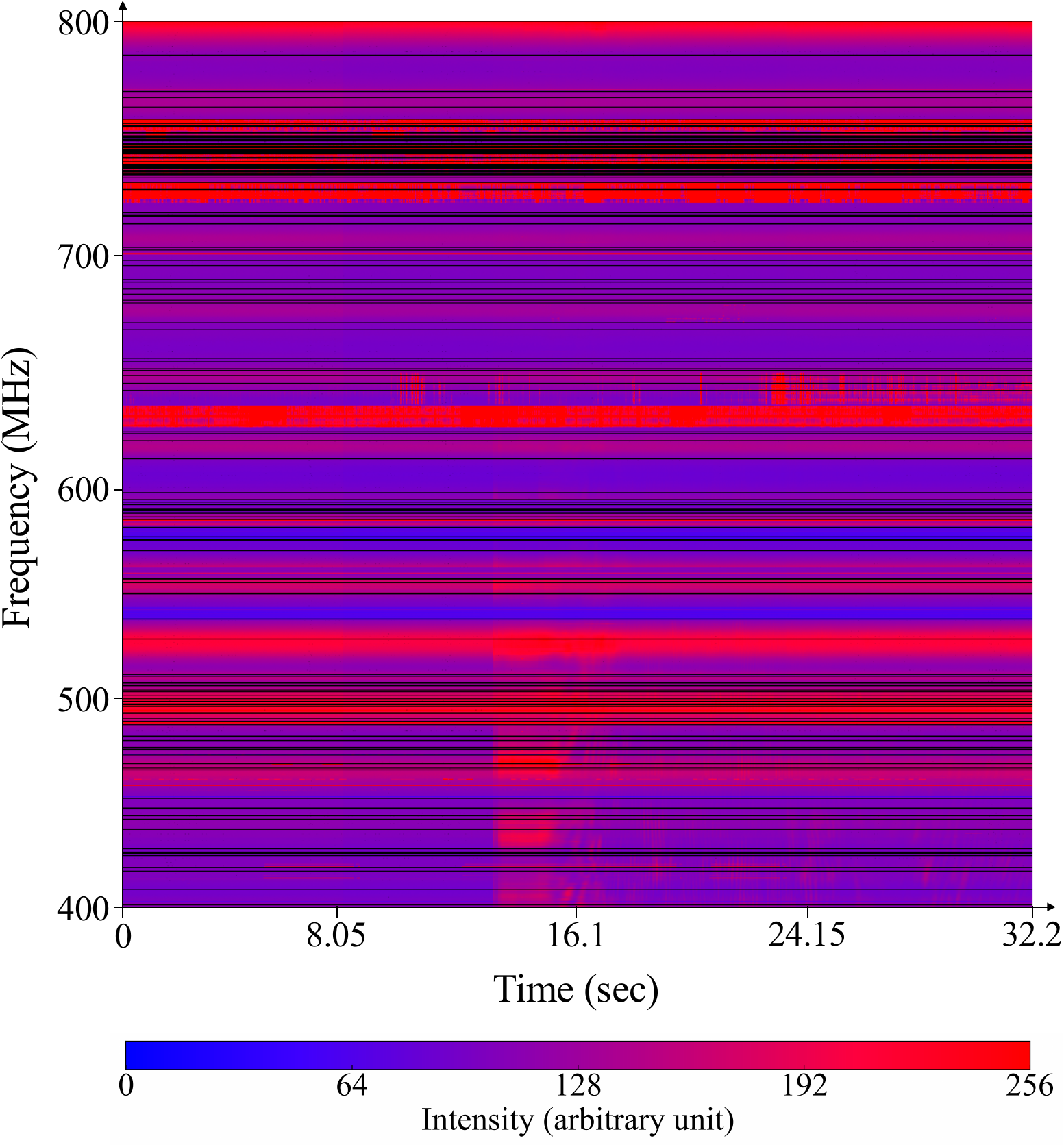}
        \hspace{0.1cm}
        \includegraphics[width=5.597150259067358cm]{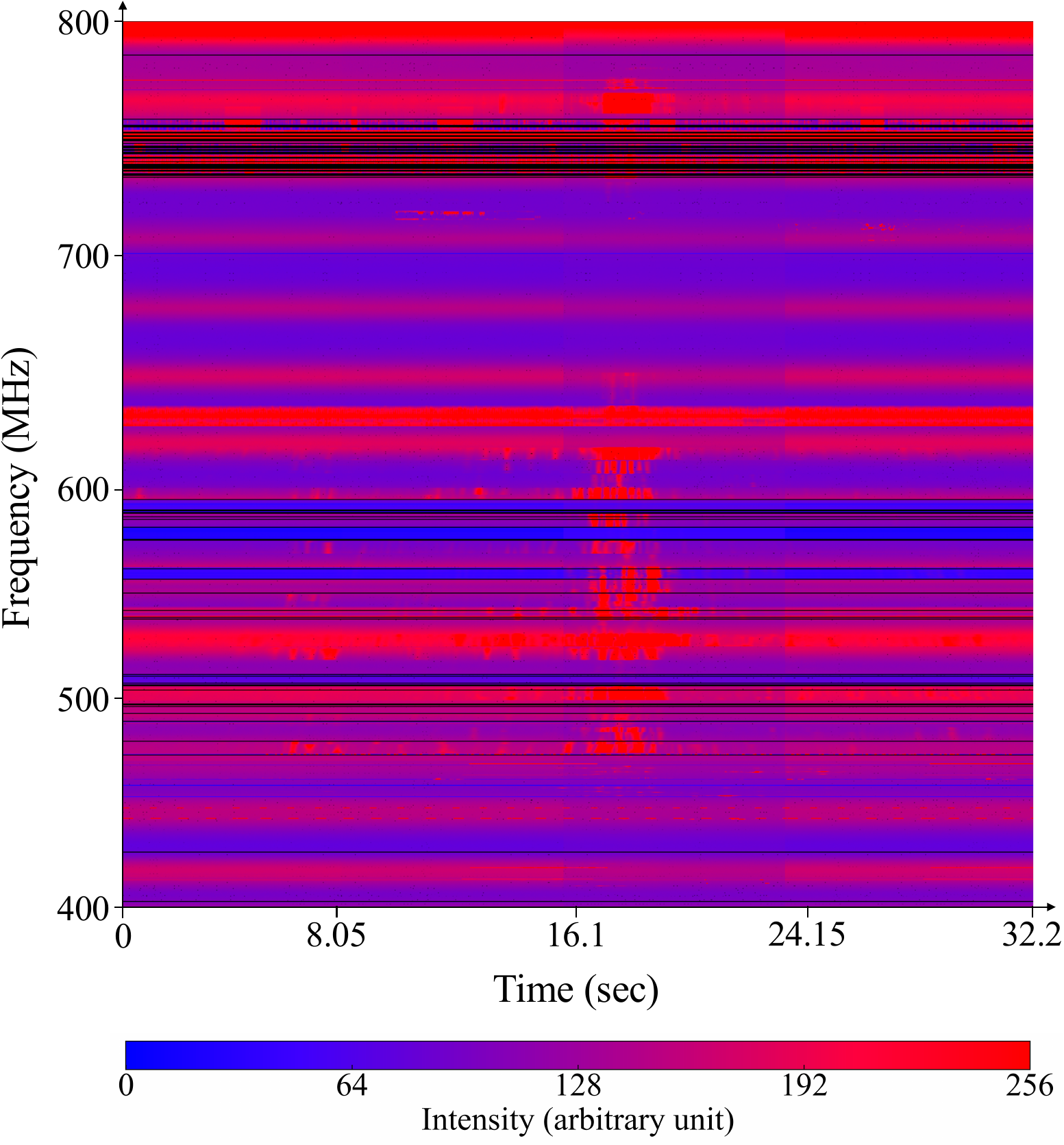}
        \hspace{0.1cm}
        \includegraphics[width=5.597150259067358cm]{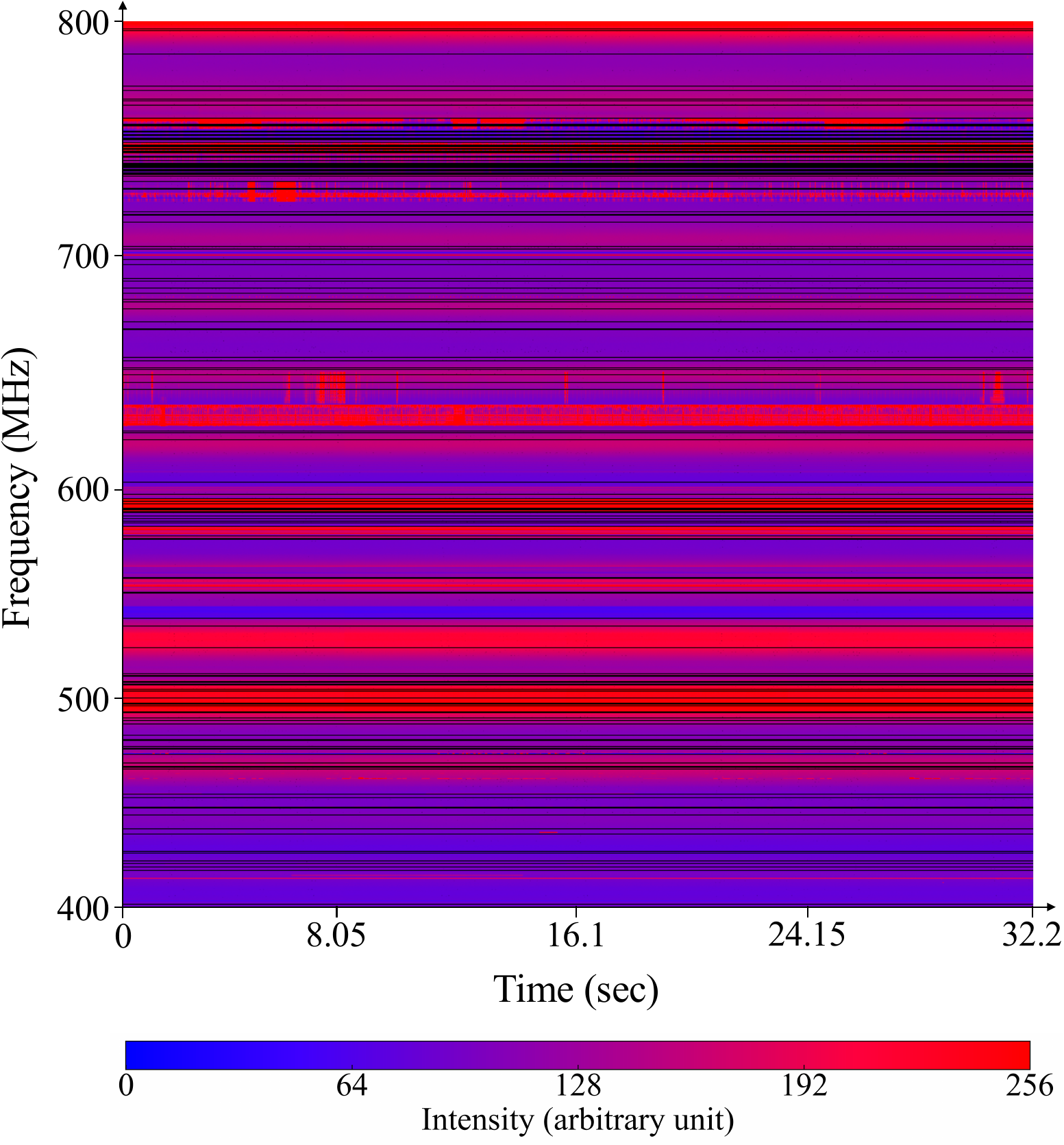}
}
\vspace{0.25cm}
\centerline{
        \includegraphics[width=5.597150259067358cm]{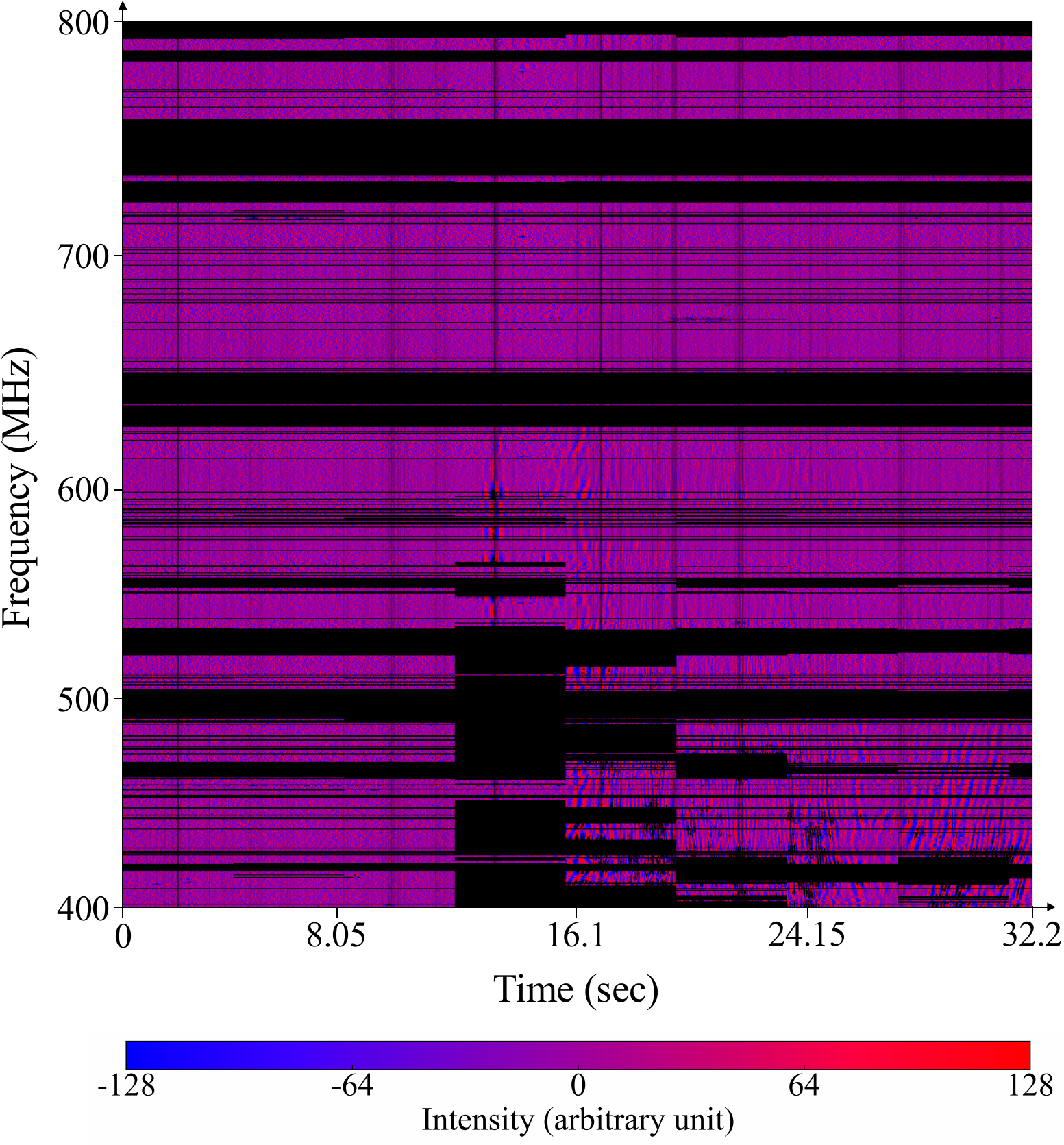}
        \hspace{0.1cm}
        \includegraphics[width=5.597150259067358cm]{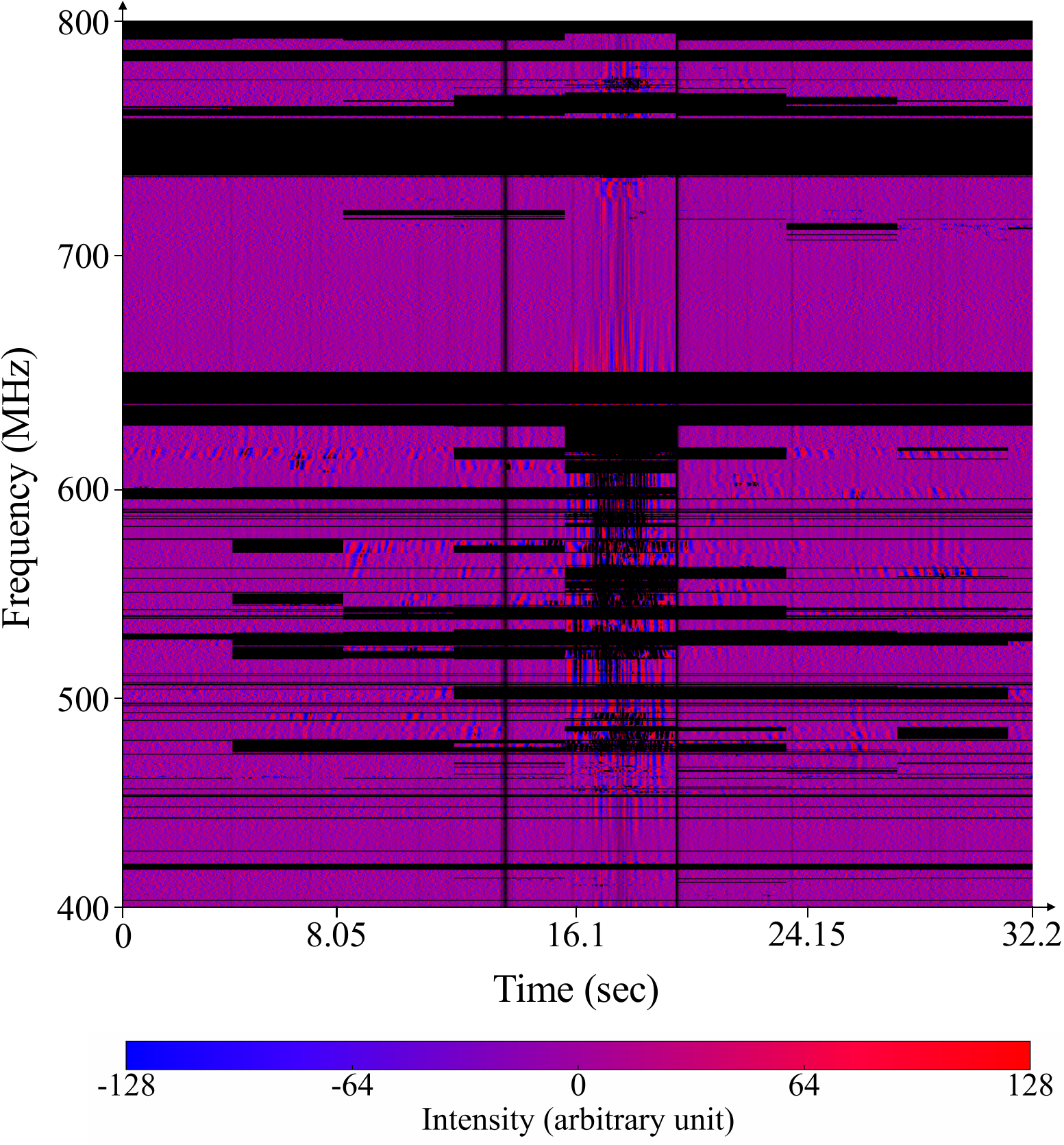}
        \hspace{0.1cm}
        \includegraphics[width=5.597150259067358cm]{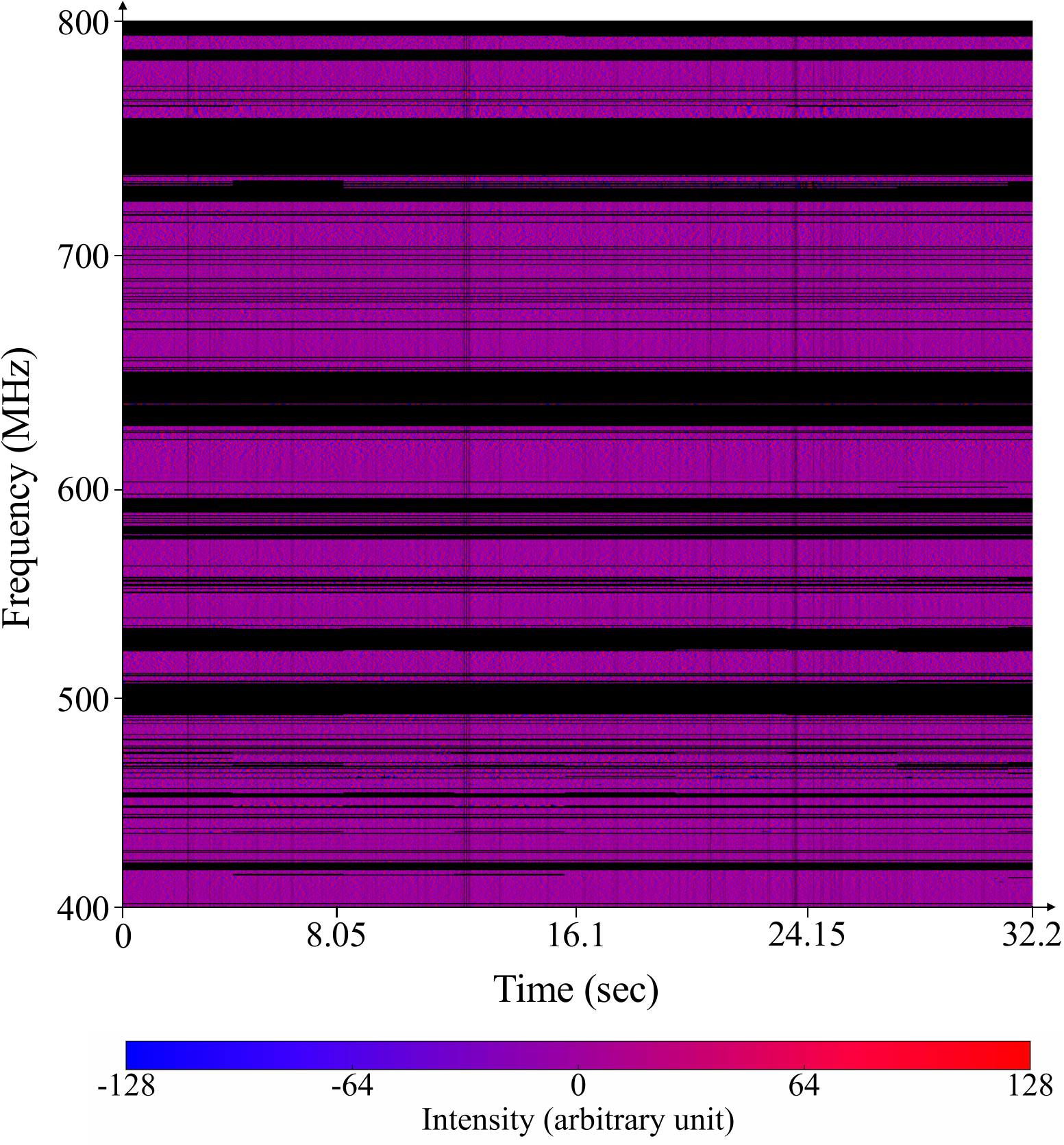}
}
\vspace{0.25cm}
\centerline{
        \hspace{-0.2cm}
        \includegraphics[width=5.8cm]{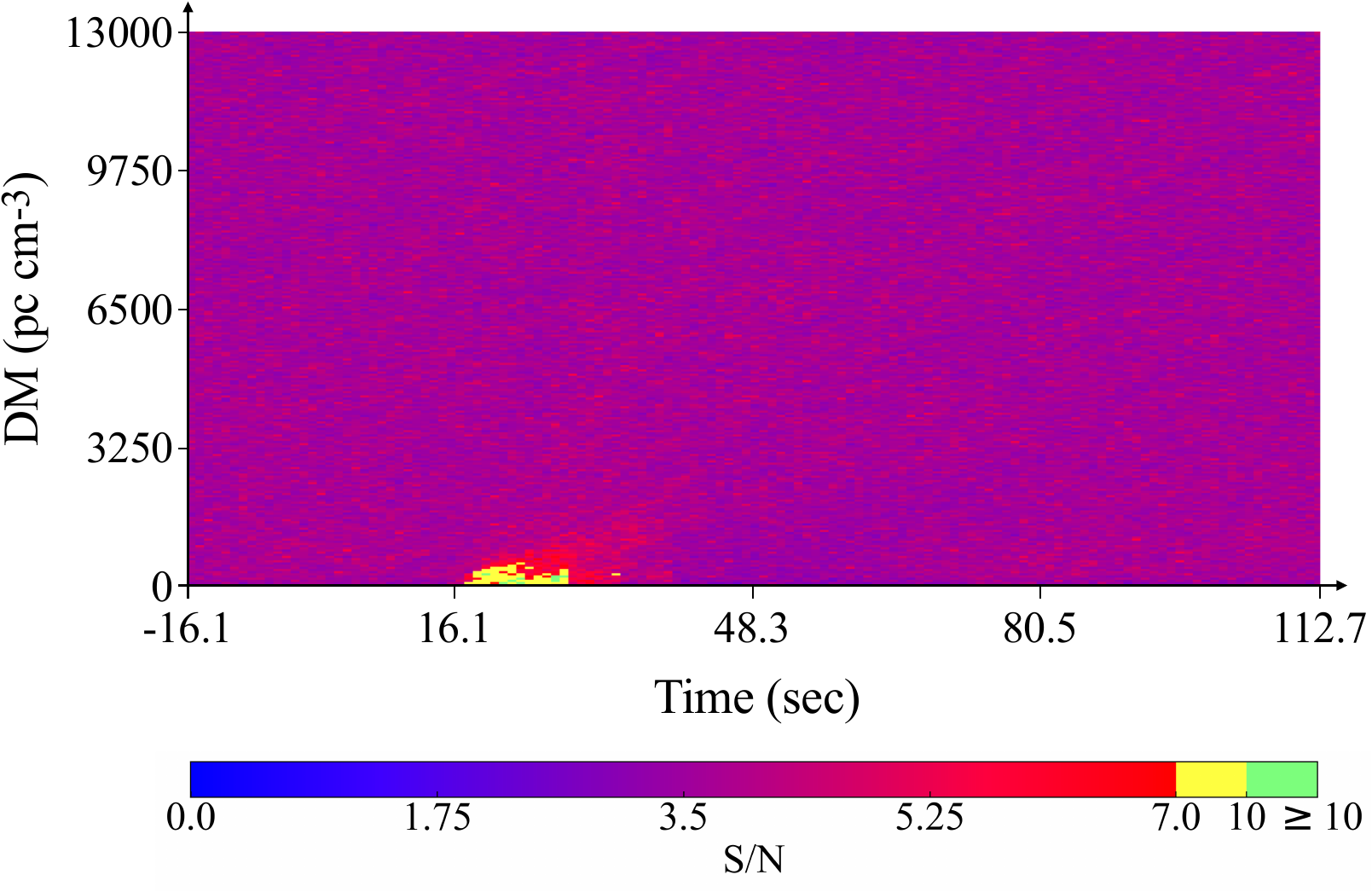}
        \includegraphics[width=5.8cm]{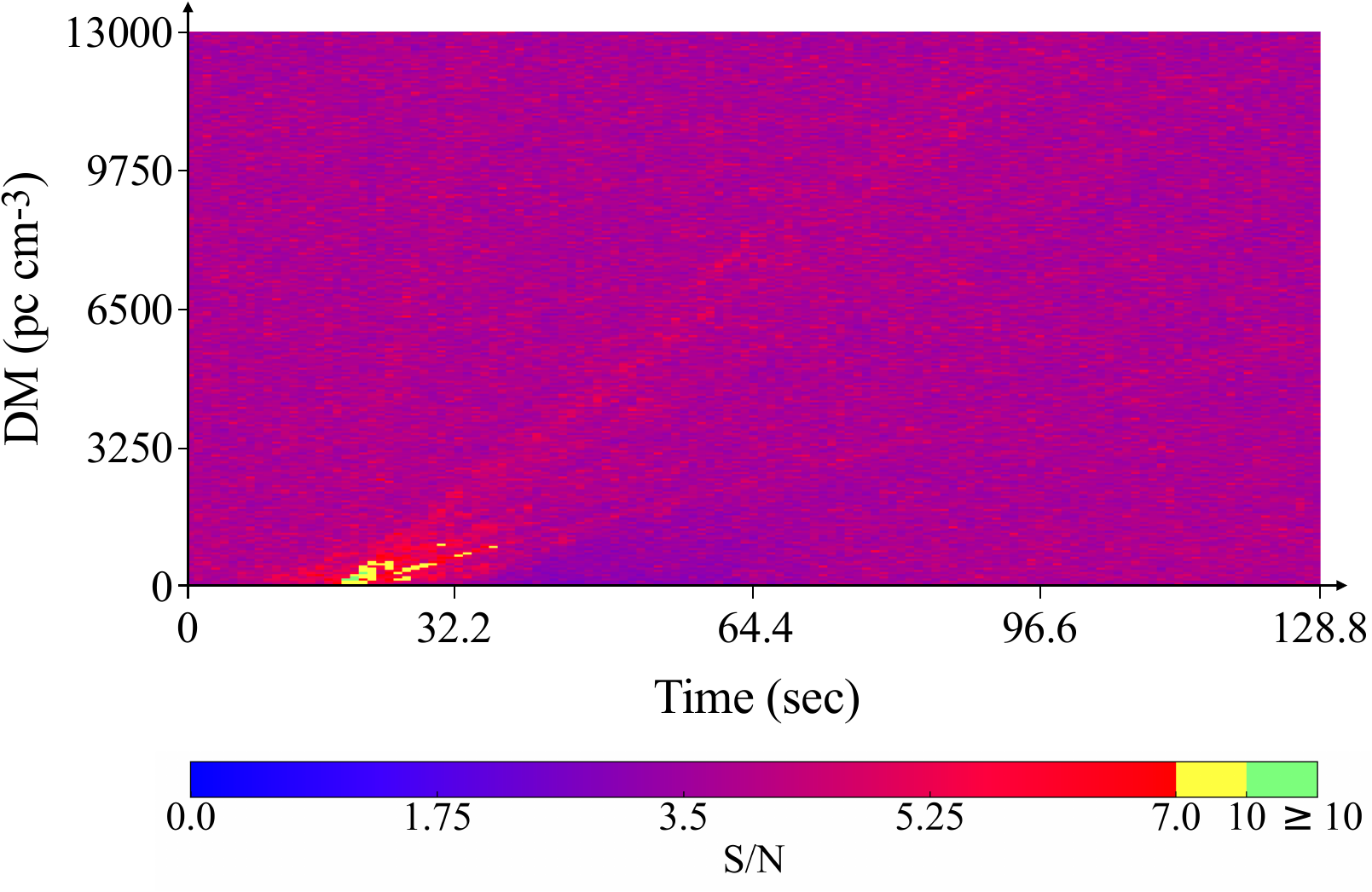}
        \includegraphics[width=5.8cm]{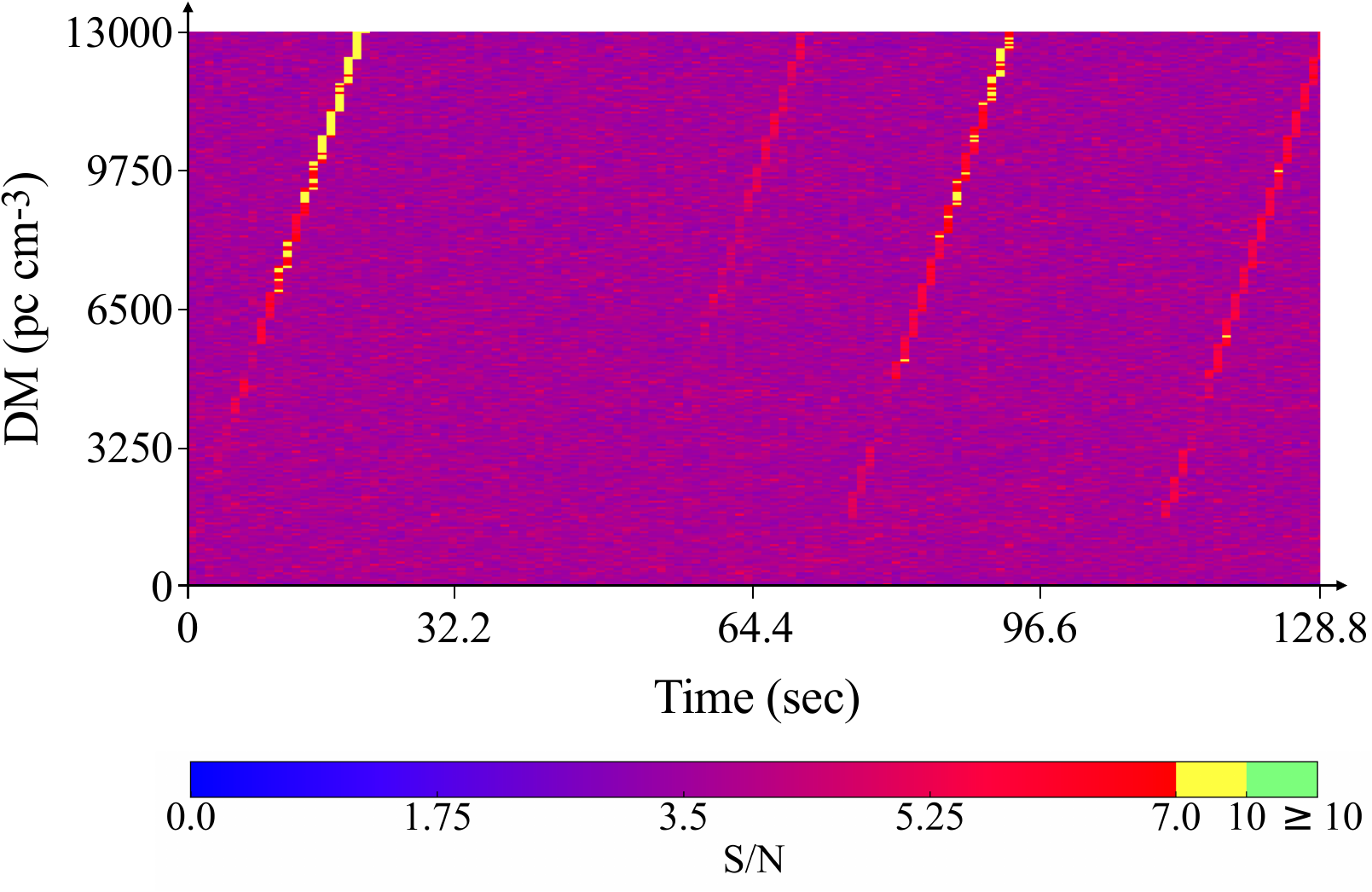}
}
\caption{CHIME/FRB L1 data for three single beams, containing impulsive RFI that potentially
         triggers false positives in the downstream pipelines.
         {\em Top row:} raw intensity with persistent features as described in
         Figure~\ref{fig:waterfall}. Black regions indicate missing data.
         {\em Middle row:} detrended intensity, including the missing data and RFI mask (black
         regions) prior to the dedispersion transform. 
         {\em Bottom row:} dedispersion space of events, including residual-induced false positives
         (left and middle panels) and subthreshold ($7 \le {\rm S/N} < 10$) high-DM false positives
         (right panel), which could result in, e.g., RFI storms (see Figure~\ref{fig:rfi_storm}).}
\label{fig:waterfall_fp}
\end{figure*}

\begin{figure*}
\centerline{
    \hspace{-0.45cm}\includegraphics[width=16.8cm]{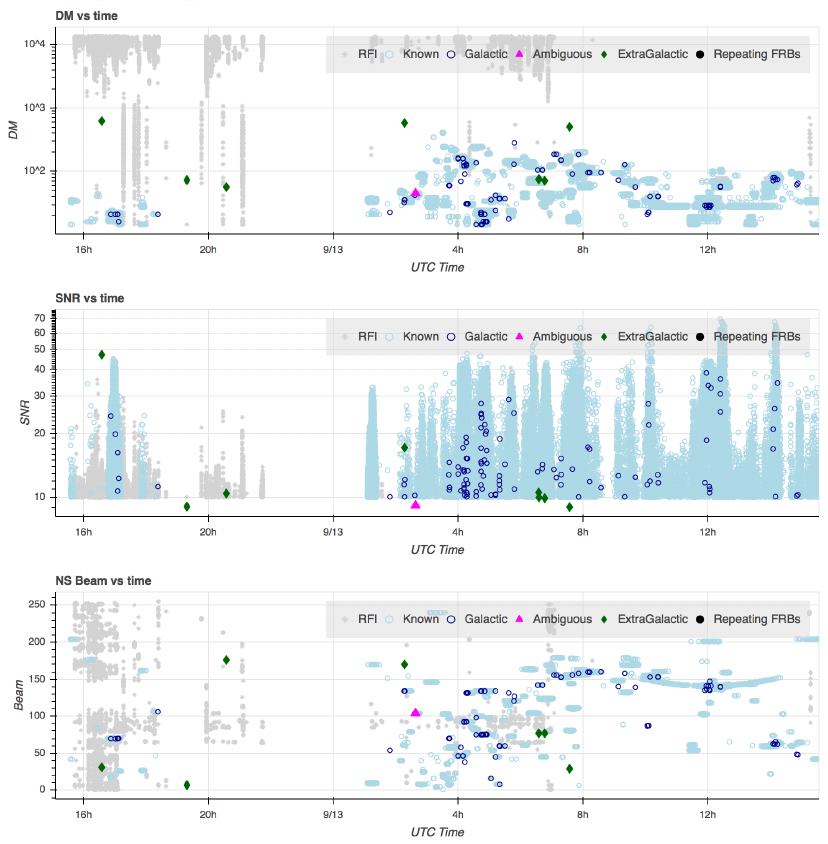}
}
    \caption{CHIME/FRB L4 events over a 24\,hr period on 2020 September 12--13 (Figure courtesy
    of the CHIME/FRB collaboration). Repeating and nonrepeating FRB candidates are marked by black
    circles and green diamonds, respectively. ``Galactic'' and ``known'' events mostly refer to the
    same set of nearby sources, such as pulsars in the Milky Way galaxy. ``Ambiguous'' events often
    require human intervention in order to be classified as Galactic or extragalactic sources. In
    this example, three ``RFI'' storms (gray points) appear at $t_1\approx16-18$\,hr (9/12),
    $t_2\approx20-22$\,hr (9/12), and $t_3\approx2-7$\,hr (9/13). {\em Top panel:} observed dispersion
    measure up to 13000 \pcc. For some unknown reason, most RFI storms tend to appear at relatively
    high DMs. In addition, aircraft transits appear as vertical lines of gray points with
    DM~$\lesssim 10^3$ \pcc at $t\approx17-21$\,hr (9/12). {\em Middle panel:} S/N
    with a minimum 8.5$\sigma$ threshold. {\em Bottom panel:} north-south sky location based on 256
    rows of digitally formed beams (each row contains four east-west beams, which are not shown here).
    The instrument has nonzero sensitivity in its sidelobes, which can continuously detect, e.g.,
    the bright pulsar PSR B0329+54 for a few hours (see the arc centered at ${\rm beam}\approx145$
    and $t\approx12$\,hr (9/13)).
    }
\label{fig:rfi_storm}
\end{figure*}

Thus far, we have described our RFI mitigation technique, which is tailored to recognize
statistical outliers within a distribution of intensity values. Assuming unlimited computational
resources, we find empirically that any changes in our standard transform chain parameters
(e.g., Table \ref{tab:clipping_transforms}) could result in either masking excessively or RFI
residuals downstream. However, the CHIME/FRB RFI environment is dynamic; unknown RFI appears
suddenly at DRAO from time to time. In particular, we have been experiencing a series of
``RFI storms'' since late-2020 (see, e.g., Figures~\ref{fig:waterfall_fp} and \ref{fig:rfi_storm}).
Such unusually intense events could escape the entire clipping transform chain, consequently
triggering $\sim$1 million false positives in a single day of operation. Although new radio
transmitters and reflective aircraft\footnote{Aircraft-triggered RFI typically escapes the L1
pipeline, causing a few thousand subthreshold ($7 \le {\rm S/N} < 10$) false positives in a day
(see, e.g., Figures~\ref{fig:waterfall_fp} and \ref{fig:rfi_storm}). We eliminate these events
in the downstream pipelines.} could cause such high numbers of false positives, the exact source of
RFI storms and their spectral properties are still an unsolved puzzle.

If highly active RFI persists for a few weeks, then we start saving intensity data to disk for
offline inspection. If the culprit is static in frequency, e.g., newly allocated radio bands,
then we update our static mask. If the culprit is dynamic, e.g., abrupt changes in the running
variance across irregular time intervals, then we might need to increase the number of clipping
iterations ($n_c$) in order to account for a period of highly active RFI. 

Another low-budget solution would be to append a few ``auxiliary'' transforms to the main RFI
chain. Table~\ref{tab:auxiliary_transforms} lists an auxiliary transform chain that we have
designed for the CHIME/FRB pipeline. The chain contains two identical transforms that operate
along the frequency axis with a relatively large downsampling factor along the time axis
($D_t=256$), potentially resulting in expanding the mask widely over the time axis. Such mask
expanding transforms can suppress RFI residuals around edges of RFI-contaminated regions that
are localized in time. On the other hand, modifying the running L1 configuration with such
coarse-grained masks could introduce a time-varying bias into the instrumental selection function.
Therefore, we generally avoid using the auxiliary transform chain (Table~\ref{tab:auxiliary_transforms})
as a hotfix in the CHIME/FRB pipeline. 

\begin{table}
\begin{center}
\begin{tabular}{ccccccccc}
\hline\hline
    $k$ &
    type &
    $\sigma_c$ &
    $\sigma_{c,{\rm in}}$ &
    $n_{c,{\rm in}}$ &
    axis &
    $N_t$ &
    $D_t$ &
    $D_\nu$ \\
    \hline
    1 & stdv & 3 & $-$ & $-$ & $\nu$ & 4096 & 256 & 2\\
    2 & stdv & 3 & $-$ & $-$ & $\nu$ & 4096 & 256 & 2\\ \hline\hline
\end{tabular}
\end{center}
    \caption{Auxiliary transform chain for enhancing the CHIME/FRB RFI mask under
    severe RFI conditions. The index $k$ specifies the position of each transform in
    the chain. The other parameters are described in Table~\ref{tab:clipping_transforms}.}
\label{tab:auxiliary_transforms}
\end{table}

\section{Discussion}
\label{sec:discussion}

The CHIME/FRB L1 pipeline is currently the leading search engine for detecting FRBs,
which are usually immersed in an influx of RFI-triggered false positives. We mitigate
RFI in a sequence of iterative operations that are fine-tuned empirically based on
statistics of intensity values in real time. In this paper, we focus on false positives
that arise in the 2D dedispersion space ($t,\DM$). However, the L1 dedispersion transform
generates a 4D trigger array indexed by (DM, scattering measure, spectral index, $t$),
presenting an opportunity for masking RFI in other dimensions.

Our RFI mitigation strategy is an attempt to minimize the overall computational cost,
false-positive and false-negative rates for the CHIME/FRB instrument. This problem could be generalized
to the continuous domain, where binary rates are replaced by ranges of parameters in the selection
function of our instrument. For instance, we could lower our false-positive rate significantly by
adding more clipping transforms with $\sigma_c < 3$ to the RFI transform chain. This would introduce
a significant bias in our sensitivity for detecting bright FRBs. In other words, any solution
to the RFI problem places a set of constraints on the instrumental selection function in the
continuous domain of FRB parameters (see, e.g., \S\ref{ssec:detrending_transform_chain}).
On the other hand, our injection system, which injects synthetic FRB signals in real time into
the pipeline allows us to characterize such biases in principle \citep{Merryfield:2022aa}.

In this work, we have presented a customized set of operations that probably does not
correspond to an optimal solution for the CHIME/FRB RFI environment. We could
in principle replace our trial-and-error strategy by an automated pipeline,
e.g., based on machine-learning techniques, that would solve the minimization problem
more accurately. Such a systematic approach would enable us to impose user-defined
constraints on the instrumental selection function for varying choices of RFI environment.
In addition, the L1 pipeline could learn from downstream pipelines through an automated
feedback mechanism; assuming that L1 could undergo a soft restart, we could modify the
transform chain on-the-fly in order to account for sudden changes in the RFI environment.
On the other hand, a time-varying transform chain might not lead to a well-characterized
instrumental selection function. We defer the study and development of these features
to future work.

Our algorithms have proven to be successful in maintaining a low false-positive rate,
while allowing for the detection of astrophysical events. Despite being effective, our
current detrenders operate independently along a single axis (time or frequency) that
may not fully capture 2D variations of RFI across the intensity plane ($t,\nu$). In
addition, detrending along the frequency axis can remove significant signal from
intrinsically wide bursts at low DMs, severely impacting the true-negative rate and
completeness \citep[see Figures 16 and 18 in][]{Collaboration:2021wz}. Furthermore,
we note that CHIME/FRB RFI mitigation and incoherent dedispersion routines were designed
based on the assumption that FRBs are isolated events in time. However, this assumption
might not be valid for repeating and multicomponent FRBs with $\lesssim 1$\,s
separation between events \citep{ch_frb_periodicity_paper}. Additionally, FRBs could
be immersed in transits of bright pulses from Galactic sources such as PSR B0329+54,
which could impact the running variance. Given the complexity of these multivariate
problems, we characterize the CHIME/FRB selection function through systematic FRB
injections into real data \citep{Collaboration:2021wz, Merryfield:2022aa}.

CHIME/FRB intensity data have provided us with a unique opportunity to investigate
a multitude of RFI-related problems that might be of interest to other experiments such
as the upcoming Hydrogen Intensity and Real-time Analysis Experiment \citep{Crichton:2022tg}
and the Canadian Hydrogen Observatory and Radio-transient Detector \citep{Vanderlinde:2019wm}.
Here, we list a few unsolved problems for future exploration:

\begin{itemize}
    \item What is the source of RFI storms? How can we suppress them without masking excessively?
    \item Most RFI storms tend to trigger false positives at high DMs (see, e.g.,
        Figure~\ref{fig:rfi_storm}). Given the CHIME/FRB L1 pipeline, would this
        be expected statistically? Could this be due to a design flaw in our RFI
        transform chain?
    \item $\sim$10\% of CHIME/FRB multibeam-detected FRB candidates seem to exhibit
        a false negative in at least one beam. Given that neighboring beams are
        expected to share the same RFI environment, what factors could contribute
        to such nondetections?
    \item Technically, we could have multiple subpipelines with different
        downsampling factors in series. Would this be useful for enhancing
        the mask over narrowband RFI?
\end{itemize}
We hope to build on this work in the future to answer some of these problems
as the CHIME/FRB instrument continues to observe the radio sky.

\acknowledgments
We acknowledge that CHIME is located on the traditional, ancestral, and unceded territory
of the Syilx/Okanagan people. We are grateful to the staff of the Dominion Radio Astrophysical
Observatory, which is operated by the National Research Council of Canada.  CHIME is funded
by a grant from the Canada Foundation for Innovation (CFI) 2012 Leading Edge Fund (Project 31170)
and by contributions from the provinces of British Columbia, Qu\'{e}bec, and Ontario. The CHIME/FRB
Project is funded by a grant from the CFI 2015 Innovation Fund (Project 33213) and by contributions
from the provinces of British Columbia and Qu\'{e}bec, and by the Dunlap Institute for Astronomy and
Astrophysics at the University of Toronto. Additional support was provided by the Canadian Institute
for Advanced Research (CIFAR), McGill University and the McGill Space Institute thanks to the Trottier
Family Foundation, and the University of British Columbia. Research at Perimeter Institute is supported
by the Government of Canada through Industry Canada and by the Province of Ontario through the Ministry
of Research \& Innovation. K.M.S. was supported by an NSERC Discovery Grant and a CIFAR fellowship.
We thank Sabrina Berger, Alice Curtin, Vicky Kaspi, Dustin Lang, Emily Petroff, Ziggy Pleunis,
and Shriharsh Tendulkar for comments on the manuscript.

\bibliographystyle{aasjournal}
\bibliography{rfi_paper.bib}

\end{document}